\documentclass[12pt,a4paper,twoside]{article}

\usepackage{latexsym}
\usepackage{epsfig}
\usepackage{psfig}
\usepackage{cite}

\newlength{\dinwidth} 
\newlength{\dinmargin}
\newcommand{\sigtot}{\sigma_{\rm tot}^{\gamma p}}
\newcommand{\pomeron}{I \! \!P}
\newcommand{\reggeon}{I \! \!R}

\begin{document}
\title {
\hspace*{-5mm}\large\rm
\bf\LARGE\boldmath Measurement of the\\ 
photon-proton total cross section at a center-of-mass
energy of 209 GeV at HERA\\
}
           

\date{~}
\author{\Large ZEUS Collaboration}

\maketitle

\vspace{-10.5cm}
\begin{flushleft}
\tt DESY-01-216 \\
December 2001 \\
\end{flushleft}
\vspace{+10cm}

\begin{abstract}
\noindent
   The photon-proton total cross section has been 
    measured in the process
   $e^+ p \rightarrow e^+ \gamma p \rightarrow e^+ X$
   with the ZEUS detector at HERA.
   Events were collected with photon virtuality $Q^2 < 0.02 \ {\rm GeV}^2$
   and average $\gamma p$ center-of-mass energy $W_{\gamma p} = 209$ GeV
   in a dedicated run, designed to control systematic effects,
   with an integrated luminosity of $49 \ {\rm nb}^{-1}$.
   The measured total cross section is
  $\sigtot = 174 \pm 1 ({\rm stat.}) \pm 13 ({\rm syst.}) \ \mu {\rm b}$.
  The energy dependence of the cross section is compatible with 
  parameterizations of high-energy $pp$ and $p\bar p$ data.
\end{abstract}
\pagestyle{plain}                   
\thispagestyle{empty}

\newpage 
\pagenumbering{Roman}
\begin{center}                                                    
{                      \Large  The ZEUS Collaboration              }                               
\end{center}                                                                                       
  S.~Chekanov,                                                                                     
  M.~Derrick,                                                                                      
  D.~Krakauer,                                                                                     
  S.~Magill,                                                                                       
  B.~Musgrave,                                                                                     
  A.~Pellegrino,                                                                                   
  J.~Repond,                                                                                       
  R.~Yoshida\\                                                                                     
 {\it Argonne National Laboratory, Argonne, Illinois 60439-4815}~$^{n}$                            
\par \filbreak                                                                                     
  M.C.K.~Mattingly \\                                                                              
 {\it Andrews University, Berrien Springs, Michigan 49104-0380}                                    
\par \filbreak                                                                                     
  P.~Antonioli,                                                                                    
  G.~Bari,                                                                                         
  M.~Basile,                                                                                       
  L.~Bellagamba,                                                                                   
  D.~Boscherini,                                                                                   
  A.~Bruni,                                                                                        
  G.~Bruni,                                                                                        
  G.~Cara~Romeo,                                                                                   
  L.~Cifarelli,                                                                                    
  F.~Cindolo,                                                                                      
  A.~Contin,                                                                                       
  M.~Corradi,                                                                                      
  S.~De~Pasquale,                                                                                  
  P.~Giusti,                                                                                       
  G.~Iacobucci,                                                                                    
  G.~Levi,                                                                                         
  A.~Margotti,                                                                                     
  T.~Massam,                                                                                       
  R.~Nania,                                                                                        
  F.~Palmonari,                                                                                    
  A.~Pesci,                                                                                        
  G.~Sartorelli,                                                                                   
  A.~Zichichi  \\                                                                                  
  {\it University and INFN Bologna, Bologna, Italy}~$^{e}$
% ...
\newpage                                                                                           
% ...
                                                           %
                                                           %
\newpage   
                                                           %
                                                           %
\begin{tabular}[h]{rp{14cm}}                                                                       
$^{a}$ &  supported by the Natural Sciences and Engineering Research                               
          Council of Canada (NSERC) \\                                                             
$^{b}$ &  supported by the German Federal Ministry for Education and                               
          Research (BMBF), under contract numbers HZ1GUA 2, HZ1GUB 0, HZ1PDA 5, HZ1VFA 5\\         
$^{c}$ &  supported by the MINERVA Gesellschaft f\"ur Forschung GmbH, the                          
          Israel Science Foundation, the U.S.-Israel Binational Science                            
          Foundation, the Israel Ministry of Science and the Benozyio Center                       
          for High Energy Physics\\                                                                
$^{d}$ &  supported by the German-Israeli Foundation, the Israel Science                           
          Foundation, and by the Israel Ministry of Science\\                                      
$^{e}$ &  supported by the Italian National Institute for Nuclear Physics (INFN) \\                
$^{f}$ &  supported by the Japanese Ministry of Education, Science and                             
          Culture (the Monbusho) and its grants for Scientific Research\\                          
$^{g}$ &  supported by the Korean Ministry of Education and Korea Science                          
          and Engineering Foundation\\                                                             
$^{h}$ &  supported by the Netherlands Foundation for Research on Matter (FOM)\\                   
$^{i}$ &  supported by the Polish State Committee for Scientific Research,                         
          grant no. 115/E-343/SPUB-M/DESY/P-03/DZ 121/2001-2002\\                                  
$^{j}$ &  partially supported by the German Federal Ministry for Education                         
          and Research (BMBF)\\                                                                    
$^{k}$ &  supported by the Fund for Fundamental Research of Russian Ministry                       
          for Science and Edu\-cation and by the German Federal Ministry for                       
          Education and Research (BMBF)\\                                                          
$^{l}$ &  supported by the Spanish Ministry of Education and Science                               
          through funds provided by CICYT\\                                                        
$^{m}$ &  supported by the Particle Physics and Astronomy Research Council, UK\\                   
$^{n}$ &  supported by the US Department of Energy\\                                               
$^{o}$ &  supported by the US National Science Foundation                                          
\end{tabular}                                                                                      
                                                           %
\newpage

\pagenumbering{arabic}
%
%
\section{Introduction}

A measurement of the photon-proton total hadronic cross section at the high
center-of-mass energy of the HERA $ep$ collider provides a test
of the asymptotic behavior of total cross sections.
The energy dependences of the
$pp$, $\bar{p}p$, $Kp$ and $\pi p$ total cross sections are successfully
parameterized by the exchanges of Regge trajectories~\cite{regge}.
Phenomenological fits~\cite{dl92,cu00,dl98,fs} based on Regge theory
are able to describe all the hadronic total cross sections in the full
energy range using the form
\begin{eqnarray}
\sigma_{\rm tot} = A \cdot s^{\epsilon} + B \cdot s^{-\eta} ,
\label{eq:regge}
\end{eqnarray}
where $s$ is the square of the center-of-mass energy and $A$ and $B$ are 
constants. The parameters
$\alpha_{\pomeron}(0)=1+\epsilon$ and $\alpha_{\reggeon}(0)=1-\eta \approx 
0.55$ denote the intercepts of the Pomeron and Reggeon
trajectories, respectively. At the highest energies, the Pomeron 
intercept describes the weak energy dependence of
hadronic total cross sections by the form
 $\sigma_{\rm tot}  \propto s^\epsilon$, where
$ 0.08 < \epsilon < 0.096$~\cite{dl92,cu00,dl98,fs}.

The soft hadronic behavior of the photon is well described by the 
vector-meson dominance model (VDM)~\cite{vmd,bauer}, in which
the photon is considered to be a superposition
of the light vector mesons ($\rho^0$, $\omega$ and $\phi$), which
 interact with the proton.
This model has been well tested at low energies~\cite{lowenergy}. 
If, at asymptotic energies, the photon can be
entirely described by the VDM, the same universal energy dependence 
is expected for the hadron-proton and 
photon-proton total cross sections.
Furthermore, if these total cross sections are dominated by the exchange
of a Pomeron trajectory at high energies, the
$pp$, $\gamma p$ and $\gamma \gamma$ total cross sections
will be related.


This paper reports a new 
ZEUS measurement of the photon-proton total cross section, 
using the reaction $e^+ p \rightarrow e^+ \gamma p \rightarrow
 e^+ X $. 
The energy of the scattered positron was measured,
thus providing a determination of the photon energy;
the limited angular acceptance for the scattered positron restricted
the virtuality of the tagged photon to very small values, 
$Q^2 < 0.02 \, {\rm GeV}^2$. This measurement requires
the accurate determination of the acceptance of the positron-tagging 
calorimeter. For this reason,
the experiment was performed under closely controlled conditions of
the positron-beam parameters,
resulting in a reduction of systematic effects. 
The detector acceptance for the hadronic final-state $X$ 
in the above reaction 
is sensitive to the topologies and relative fractions 
of the photoproduction subprocesses.
For the previous ZEUS measurements~\cite{zeusold}, 
the subprocess cross sections were extracted through
fits to the distribution of the energy deposits
in the main ZEUS calorimeter.
Detailed studies of several of the subprocesses have subsequently been 
published by ZEUS~\cite{rho,omega,phi,pdisselasticratio,gdissfrac},
permitting improved estimates of the respective cross sections.  
In the present paper, these subprocess measurements are used as constraints to
obtain a more accurate measurement of the  photoproduction total
cross section than was previously possible.                                
Because of these improvements, the results of this 
paper supersede those of the previous publications~\cite{zeusold}. 
The photon-proton total cross section has also been measured
at a similar center-of-mass energy by the H1 Collaboration~\cite{h1new}.

\section{Kinematics}

The photon-proton total cross section has been measured
in the process 
$e^+p \rightarrow e^+ \gamma p \rightarrow e^+ X$,
where the interacting photon is almost real.
The event kinematics may be described in terms of Lorentz-invariant
variables: the photon virtuality, $Q^2$, and the event inelasticity, $y$,
defined by
\begin{displaymath}
  Q^{2} = -q^{2} = -(k - k^{\prime})^2 \, 
  \nonumber
\end{displaymath}
and
\begin{displaymath}
  y = \frac{p\cdot q}{p\cdot k} \, ,
  \nonumber
\end{displaymath}
where $k$, $k^{\prime}$ and $p$ are the four-momenta of the
incoming positron, scattered positron and incident proton, respectively.
The square of the photon-proton center-of-mass energy is given by
\begin{displaymath}
  W^{2}_{\gamma p} = ( q + p )^{2} \, .
  \nonumber
\end{displaymath}
These variables can be expressed in terms of the experimentally 
measured quantity $E_e^{\prime}$ using
\begin{displaymath}
  Q^2 = Q^2_{\rm min} + 4 E_e E_e^{\prime} \sin^2\frac{\vartheta}{2} \, ,
  \nonumber
\end{displaymath}
\begin{displaymath}
  y = 1 - \frac{E_e^{\prime}}{E_e} \cos^2\frac{\vartheta}{2}  
    \simeq 1 - \frac{E_e^{\prime}}{E_e} \, ,
  \nonumber
\end{displaymath}
\begin{displaymath}
 W_{\gamma p} = 2 \sqrt{E_e E_p y} \, ,
 \nonumber
\end{displaymath}
where
\begin{displaymath}
  Q^{2}_{\rm min} = \frac{m^{2}_{e}y^{2}}{1 - y} \, ,
  \nonumber
\end{displaymath}
$E_{e}$, $E_{e}^{\prime}$ and $E_{p}$ are the energies of the 
incoming positron, scattered positron and incident proton, respectively,
$\vartheta$ is the positron scattering angle with respect
to the initial positron direction and $m_e$ is the positron mass.
The scattered positron was recorded in a positron tagger close to the beam
line, restricting the production angle, $\vartheta$ 
 (and hence $Q^2$), to small values.
The photon virtuality ranged from the kinematic minimum,
$Q^{2}_{\rm min} \simeq 10^{-7}\ {\rm GeV}^2$, up to 
$Q^2_{\rm max} \simeq 0.02 \ {\rm GeV}^2$,
given by the acceptance of the positron tagger. The median
$Q^2$ is about $5 \times 10^{-5}$ GeV$^2$.  

The equivalent photon approximation~\cite{gribov} relates
 the electroproduction cross section to
a photoproduction cross section.
The double-differential $ep$ cross section can be written as
\begin{eqnarray}
\frac{d^{2}\sigma^{ep}_{\rm tot}(y,Q^{2})}{dydQ^{2}}
  =\left.\frac{\alpha}{2\pi}\frac{
1}{Q^{2}}\right[ \
\left( \frac{1+(1-y)^{2}}{y}
      -\frac{2(1-y)}{y} \frac{Q^{2}_{\rm min}}{Q^{2}} \right)
  \ \cdot \sigma^{\gamma p}_{T}(y,Q^{2}) \hspace*{0.2cm} \nonumber \\
+ \left.\frac{2(1-y)}{y} \cdot \sigma^{\gamma p}_{L}(y,Q^{2})\right] \, ,
\label{eqn:epdoubdiff}
\end{eqnarray}
where $\sigma^{\gamma p}_{T}$ is the cross section for interactions between
the proton and a photon with transverse polarization, and
$\sigma^{\gamma p}_{L}$ is the cross section for interactions
with longitudinally polarized photons.
The longitudinal cross section is expected to be small
($\sigma^{\gamma p}_{L} / \sigma^{\gamma p}_{T} < 0.1\%$~\cite{lohrmann}), 
and has been ignored, as has the $Q^2$ dependence of $ \sigma^{\gamma p}_{T}$
($<$0.1\% in the range of $Q^2$ of this measurement~\cite{vmd,bauer}).

Integrating Eq.~(\ref{eqn:epdoubdiff}) over $Q^2$ 
gives the single $ep$
differential cross section in terms of the $\gamma p$ total cross section:
\begin{equation}
  \frac{d\sigma^{ep}_{\rm tot}(y)}{dy}=
   \frac{\alpha}{2\pi}
  \left[ \frac{1+(1-y)^{2}}{y}\ln\frac{Q^{2}_{\rm max}}{Q^{2}_{\rm min}}
     -\frac{2(1-y)}{y} \left( 1-\frac{Q^{2}_{\rm min}}{Q^{2}_{\rm max}} \right)
  \right]
  \sigtot(y) =F_{\gamma}(y)\sigtot(y) \, .
\label{eq:singlediff}
\end{equation}
Integrating Eq.~(\ref{eq:singlediff}) between $y_1$ and $y_2$, 
and neglecting the 
weak $y$-dependence of $\sigtot(y)$ in the integral, the electroproduction
cross section is
\begin{displaymath}
\sigma_{\rm tot}^{ep}=\int_{y_1}^{y_2}F_{\gamma}(y)
\sigtot(y) \, dy = f_{\gamma} \sigtot \, , \nonumber
  \nonumber
\end{displaymath}
where 
\begin{eqnarray}
f_{\gamma} = \int_{y_1}^{y_2}F_{\gamma}(y) dy \, ,
\label{eq:flux}
\end{eqnarray}
and $y_1=0.42$ and $y_2=0.56$ are derived from the minimum and maximum
detected positron energies, respectively (see Section~\ref{sec:expt_cond}).

\section{Experimental conditions}
\label{sec:expt_cond}

The data were taken with the ZEUS
detector during a dedicated run in 1996, when HERA
collided 820 GeV protons with 27.5 GeV positrons.
The proton and positron beams each contained 177 colliding bunches, 
together with 3 additional unpaired proton and 31 unpaired positron 
bunches.  
These additional bunches were used for background studies.  
The time between bunch crossings was 96 ns.

The ZEUS detector has been described in detail elsewhere~\cite{zeusdet}.
Most important for this measurement are the uranium-scintillator
sampling calorimeter (CAL)~\cite{cal}, 
the central tracking detector (CTD)~\cite{ctd}
and two lead-scintillator calorimeters close to the $e^+$ beampipe 
at $Z=-107$~m (photon tagger) and $Z=-35$~m (positron or ``35m" 
tagger)~\cite{lumi}.
The CAL is separated into three parts, 
forward{\footnote{%
The ZEUS coordinate system is a right-handed Cartesian system, with the $Z$
axis pointing in the proton beam direction, referred to as the ``forward
direction'', and the $X$ axis pointing left towards the center of HERA.
The coordinate origin is at the nominal interaction point.
The pseudorapidity is defined as 
$ \eta = -{\ln} \left( {\tan} \frac{\theta}{2} \right) $,
where the polar angle, $\theta$, is measured with respect to the proton
beam direction.  The azimuthal angle is denoted by $\phi$. 
}
(FCAL, $2.6^{\circ} < \theta < 36.7^{\circ}$),
barrel (BCAL, $36.7^{\circ} < \theta < 129.1^{\circ}$),
and rear (RCAL, $129.1^{\circ} < \theta < 176.2^{\circ}$).  
Each CAL part is longitudinally segmented into 
electromagnetic (EMC) and
hadronic (HAC) sections.  Each section is further subdivided 
transversely into cells
of typically $5 \times 20 \ {\rm cm}^2$ 
($10 \times 20 \ {\rm cm}^2$ in RCAL) for the EMC and 
$20 \times 20 \ {\rm cm}^2$ for the HAC sections.
The total number of CAL cells is 5918.
The CAL relative energy resolution for
electromagnetic showers is $\sigma_E/E \simeq 0.18 /\sqrt{E({\rm GeV})}$
and for hadronic showers is $\sigma_E/E \simeq 0.35 /\sqrt{E({\rm GeV})}$
under test-beam conditions.  The CAL also provides
timing information, with a resolution of 1 ns for energy deposits
greater than 4.5 GeV.
The CTD operates in a 1.43 T solenoidal magnetic field,
and has a relative transverse-momentum resolution for full-length tracks of
$\sigma_{p_T}/p_T = 0.0058 \, p_T  \oplus 0.0065 \oplus 0.0014/p_T$,
with $p_T$ measured in GeV.
The photon-tagger relative energy resolution is
$\sigma_E / E = 0.23 / \sqrt{E({\rm GeV})}$, and
the 35m-tagger relative energy resolution is
$\sigma_E / E = 0.19 / \sqrt{E({\rm GeV})}$.
Both the photon tagger and the 35m tagger are also equipped with
shower-position detectors. These consist of both
horizontal and vertical scintillator strips 
providing a position resolution of 0.3 cm in both $X$ and $Y$.

The luminosity was measured via the Bethe-Heitler bremsstrahlung process 
$ep\to e\gamma p$, using the photon tagger~\cite{lumi},
which detects photons
with scattering angles smaller than 
$\vartheta = 180^{\circ} - \theta = 0.54$ mrad.  
The accumulated luminosity used for this measurement was
49.26 $\pm$ 0.54 (syst.) nb$^{-1}$. 

For this measurement, 
the hadronic final-state $X$ was detected in the main ZEUS
detector and 
the final-state positrons were tagged in the 35m tagger
after traversing a window in the beam pipe.
The geometric acceptance of the 35m tagger restricts the kinematic
range of the detected positrons to approximately
$5 < E'_e < 20 \ {\rm GeV}$ and $Q^2 < 0.02 \ {\rm GeV}^2$.
The calculation of the geometric acceptance of the 35m tagger is complicated
because the scattered
positrons detected in the 35m tagger traverse the positron
beampipe after passing through two bending magnets and three
quadrupoles.  Furthermore, scattered positrons have lower energy
than the beam positrons and progressively leave the magnetic axis
and are subject, in particular, to quadrupole fringe fields.
Particular care, therefore, was taken to tune the beamline simulation
before calculating this acceptance.

During normal HERA running, the positron-beam position and
tilt at the interaction point vary over time.
The positron-beam tilt may be monitored using the position of 
bremsstrahlung photons detected in the photon tagger.
For this dedicated run, the beam position and tilt were controlled,
carefully monitored and found to be very stable.

Bremsstrahlung data were taken immediately preceding the
primary photoproduction data run to determine the effect,
on the 35m-tagger bremsstrahlung 
acceptance, of variations in the positron-beam tilt and to tune 
beamline and detector simulations.
From these data, the 35m-tagger acceptance was found
to depend on the horizontal ($X$) tilt, but not significantly on the 
vertical ($Y$) tilt. 
The positron-energy interval of $12<E'_e<16 \ {\rm GeV}$ was
chosen for the $\sigtot$ measurement as the range least sensitive to
variations of the positron-beam tilt.  From this range of selected positron
energies, the photon flux factor as determined from Eq.(\ref{eq:flux}) 
is $f_{\gamma}=0.004916$.

\section{Event selection and background subtraction}

Events were selected online by the three-level
trigger system of ZEUS.  
The CAL is segmented into trigger towers~\cite{caltrigger}
that are approximately projective;
the towers consist of an EMC and a HAC part.
In RCAL, the main component used in the CAL trigger,
the EMC section of a typical trigger tower consists of 
two cells. The trigger required a measured
energy deposit of more than 5 GeV  in the 35m tagger
in coincidence with 
a summed energy deposit in the RCAL EMC trigger towers of either more 
than 464 MeV (excluding the 8 towers immediately adjacent to the beampipe) 
or 1250 MeV (including those towers).
In addition, the timing information from the CAL
was required to be consistent with an $ep$ collision. 

The offline event-selection cuts on the 35m tagger and
reconstructed RCAL energies were tighter than those
applied at the trigger level.  A positron energy in the
range $12<E'_e<16 \ {\rm GeV}$ (see Section~\ref{sec:expt_cond})
was required in the 35m tagger.
The energy requirement in the RCAL EMC section, summed over all cells above
threshold, was: either more than 
600 MeV (excluding the trigger towers immediately adjacent to the beampipe);
more than 1550 MeV (including those towers); or 
the sum of any two trigger towers was more than 850 MeV.

The CAL trigger-tower energies for every event were readout~\cite{fastclear},
permitting detailed trigger-efficiency studies.
The offline energy thresholds were found to be well above the region of
low-energy trigger 
inefficiencies, so that the resulting event selection was unaffected by the 
trigger cuts.  

Positron-proton and positron-gas bremsstrahlung events in coincidence with 
RCAL energy deposition comprise the 
largest backgrounds in the online sample.  
Most of these background
events were rejected offline by a cut on the photon-tagger energy,
$E_{\gamma}<1 \ {\rm GeV}$.  
The residual $(1.26 \pm 0.26)\%$ positron-gas bremsstrahlung background 
in the sample was estimated using 
events associated with unpaired positron bunches, and was  
statistically subtracted from the photoproduction 
distributions used to calculate the cross section.  The remaining
number of events is $N = 22533 \pm 162$.

The background from proton-gas collisions, as measured using proton-only
bunches, was found to be negligible. 
The final event sample was corrected for two background
effects:
\begin{itemize}
\item a $0.99 \pm 0.01$ correction factor was applied 
to remove bremsstrahlung events which remained in the sample
owing to the $97.0\%$ acceptance of the photon tagger;
\item a $1.043 \pm 0.002$ factor was applied to
correct for photoproduction events lost due
to overlays with bremsstrahlung events; the probability of overlay events 
was estimated using events triggered from random $ep$ bunch crossings.

\end{itemize}

\section{Simulation of photoproduction processes}
\label{sec:mc}

The various physical processes that contribute to the hadronic total cross
section are characterized by different distributions in energy and angle
of the particles in the final state.
To determine the acceptance of the CAL (see Section 6)
for the various contributing 
processes, the simulation of the photoproduction sample was separated 
into the following subprocesses: 
\begin{itemize}
\item elastic: $\gamma p \rightarrow V p$, where
      $V$ is one of the light vector mesons $\rho^0$, $\omega$ or $\phi$;
\item proton dissociative:  $\gamma p \rightarrow V N$, where $N$ is a
      hadronic state  into which the proton diffractively dissociates;
\item photon dissociative: $\gamma p \rightarrow G p$, where $G$ is a 
      hadronic state  into which the photon diffractively dissociates;
\item double dissociative: $\gamma p \rightarrow G N$; 
\item hard non-diffractive: $\gamma p \rightarrow X$;
\item soft non-diffractive: $\gamma p \rightarrow X$. 
\end{itemize}
The first four processes are diffractive, in the sense that they can
be parameterized at high energies 
in Regge theory by the exchange of a Pomeron trajectory.
The hard non-diffractive part of photoproduction 
consists, in leading-order QCD, of direct and resolved photon components, 
which can be calculated perturbatively.
The largest contribution to the cross section comes from the soft 
non-diffractive process.  
Two independent Monte Carlo (MC) samples were generated to simulate the various
hadronic final states for photoproduction.
The first sample was generated using PYTHIA 5.7~\cite{pythia}, with radiative
 corrections
calculated by HERACLES 4.6~\cite{heracles}, and events from 
each of the subprocesses were selected separately.
For the second sample, HERWIG 5.9~\cite{herwig} was used for the 
non-diffractive 
reactions, while the diffractive processes were again 
generated with PYTHIA. Hard non-diffractive
photoproduction events were simulated in both MC generators at leading
order with parton showers, using CTEQ4L~\cite{cteq} 
and GRVG LO~\cite{grv} for the proton and photon parton distributions,
respectively; a minimum transverse momentum
of the partonic hard scatter, $\hat{p}^{\rm min}_T$ = 2.5 GeV, was used.
These two samples were passed through the trigger and detector simulations 
and offline analysis.

\section{Acceptance of the hadronic final state}
\label{sec:calacc}

\begin{table}
\centering
\vspace*{0.1cm}
\begin{tabular}{|ll|c|c|}
\hline
Process & & $A_{\rm CAL}$ & Fraction  \\
\hline
elastic              
   & $\gamma p \rightarrow V p$           & $0.477 \pm 0.009$ & 0.091  \\
proton dissociative  
   & $\gamma p \rightarrow V N$           & $0.531 \pm 0.012$ & 0.045\\
photon dissociative  
   & $\gamma p \rightarrow G  p$  & $0.803 \pm 0.006$ & 0.133 \\
double dissociative  
   & $\gamma p \rightarrow G N$  & $0.824 \pm 0.007$  & 0.065\\
hard non-diffractive 
   & $\gamma p \rightarrow X $   & $0.858 \pm 0.005$ & $0.166\pm 0.019$ \\
soft non-diffractive 
   & $\gamma p \rightarrow X $   & $0.832 \pm 0.003$ & $0.498\pm 0.058$\\
\hline
\end{tabular}
\caption{CAL acceptance and fractions 
for the various photoproduction subprocesses 
for PYTHIA. Those fractions that have no listed uncertainty were fixed in
the fitting procedure.}
\label{table:acc}
\end{table}

To find the overall acceptance of the CAL,
a weighted sum of the MC photoproduction subprocesses was simultaneously
fitted to the invariant mass of the hadronic final state detected in the CAL,
$M_X^{\rm CAL}$, and the
number of CAL cells, $N_{\rm cells}$, in the data.
The CAL acceptance for each subprocess was calculated as the
fraction of generated events that pass the RCAL offline cuts;
these individual acceptances are shown in Table~\ref{table:acc}.
The overall acceptance of the CAL was then calculated using
the fitted fraction and the acceptance for each subprocess.
Because the data distributions were not described perfectly by the MC,
the distribution of the soft 
non-diffractive subprocess, which is the MC process least constrained by 
experiment, was re-weighted. 
The re-weighting function was calculated separately for each of the following
distributions:   
the pseudorapidity of the CAL cells, the total transverse energy and the 
RCAL energy.
 
To determine the subprocess fractions,
the direct-to-resolved cross-section ratio was fixed to that used in the 
PYTHIA MC generator and
factorization was assumed to estimate the double dissociative cross section
by using
\begin{displaymath}
\frac{\textstyle{\sigma (\gamma p \rightarrow V p)}}
     {\textstyle{\sigma (\gamma p \rightarrow V N)}}
=
\frac{\textstyle{\sigma (\gamma p \rightarrow G p)}}
     {\textstyle{\sigma (\gamma p \rightarrow G N)}}. \nonumber
\end{displaymath}

The elastic $\rho^0$~\cite{rho}, $\omega$~\cite{omega} 
and $\phi$~\cite{phi} cross sections, 
the elastic to proton-dissociative ratio~\cite{pdisselasticratio},
and the photon-dissociative fraction~\cite{gdissfrac} were all fixed 
to values obtained from ZEUS measurements extrapolated 
to the $W_{\gamma p}$ of the present measurement. 
This reduced the fit parameters to two: the soft and hard non-diffractive 
fractions. 

The data and fitted MC distributions of $M_X^{\rm CAL}$ and $N_{\rm cells}$
are shown in 
Figs.~\ref{fig:fig2mxlin} and~\ref{fig:fig2ncellin}, respectively. 
The subprocess contributions are also shown.
The detailed description of the systematic uncertainty calculation,
shown as a band in the figures, is given in Section~\ref{sec:systs}.
As a consistency check, the RCAL energy distribution, which is not used 
in the fit, is shown in Fig.~\ref{fig:fig2rcallog}
with the MC distributions.
In Fig.~\ref{fig:fig2rest}, a comparison of the data and MC distributions 
is shown for various CAL quantities.
Overall, the data are well described by the fitted MC samples.
The CAL acceptance is
$A_{\rm CAL} = 0.781^{+0.022}_{-0.016} ({\rm syst.})$.
The largest contribution to the systematic uncertainty 
results from the different MC models
used to calculate the acceptance and is discussed in Section~\ref{sec:systs}.

\section{Acceptance of the 35m tagger}
\label{sec:tagacc}

The acceptance of the 35m tagger for scattered positrons was determined from
a PYTHIA MC simulation of photoproduction.  The
bremsstrahlung data were used to determine the parameters of the incoming
positron beam to be used in the BREMGE MC generator~\cite{bremmc}. 

For the range of bremsstrahlung photon and positron energies 
relevant to the current measurement, bremsstrahlung photons were detected
in the photon tagger and bremsstrahlung positrons in the 35m tagger
with high efficiency.  In particular, the photon-tagger acceptance
for photons with energy greater than 1 GeV
was $97.0\%$.
The 35m-tagger acceptance for scattered bremsstrahlung positrons 
was defined to be the number
of events containing a positron with $E'_e>5 \ {\rm GeV}$ in the 35m tagger
and a photon with energy $E_{\gamma}>5 \ {\rm GeV}$ in the photon detector,
divided by the total number of events
containing a photon with energy $E_{\gamma}>5 \ {\rm GeV}$
in the photon detector.
Because the 35m-tagger bremsstrahlung acceptance was found
to be sensitive primarily to variations in the horizontal plane,
the MC alignment tuning was restricted to the $X$-vertex
position and $X$ tilt.
Four distributions of both bremsstrahlung data and 
bremsstrahlung MC~\cite{bremmc} events were used to
form a $\chi^2$, which was minimized with respect to the MC $X$-vertex
position and $X$ tilt:  the photon energy; the positron energy;
the 35m-tagger bremsstrahlung acceptance;
and the average positron $X$ position versus energy measurement.
The bremsstrahlung data and MC distributions after tuning are
shown in Fig.~\ref{fig:bsopt}.
There is good agreement between data and MC events except in 
Fig.~\ref{fig:bsopt}c) for photon energies above $\sim$15 GeV. This
region is not used in the current measurement. 

The tuned values were used in the
generation of PYTHIA MC samples and the 35m-tagger photoproduction 
acceptance was calculated.
In Fig.~\ref{fig:gpopt}, the measured and simulated energy spectra of
the scattered 
positron are compared.  Figure~\ref{fig:gpcorr} shows 
the correlation between the scattered-positron 
position and energy, as measured with 
the 35m tagger, for the data and Monte Carlo simulation.
The agreement between the data and the PYTHIA events is good.
The resulting 35m-tagger photoproduction acceptance for
$12 < E'_e < 16$ GeV is
  $A_{\rm 35m} = 0.693  \pm 0.050 ({\rm syst.})$.

\section{Radiative corrections}
%
The positron that initiates the $\gamma p$ interaction is subject to
QED radiation in both the initial and final states. This changes the
kinematics of the reaction and hence the measured cross section. 
The effect of QED radiation on the measurement can be greatly reduced
by excluding hard initial-state bremsstrahlung. 
This was achieved by a veto on photons with energy larger than 1 GeV 
in the photon detector.
The influence of radiation on the measured cross section can be 
described by a correction factor which is the ratio
of the Born cross section to that including QED radiation.

The calculation of the correction factor was carried out with the
HERACLES MC program that includes the positron-beam angular 
divergence at the interaction point. The result for the correction
factor is $ 0.981 \pm 0.007 ({\rm syst.})$.

The systematic uncertainties on the correction factor were estimated 
from the range of values
 obtained for different parameterizations of the cross section, from a
 comparison with the analytic calculation of HECTOR 1.11~\cite{hector}
and from varying the values of the photon-energy cut and angular acceptance
 within the experimental uncertainties.

\section{Systematic uncertainty estimation}
\label{sec:systs}

The systematic uncertainties in the $\sigtot$ measurement
come primarily from the uncertainties in the
determination of the CAL and 35m-tagger acceptances.

The systematic uncertainty on the CAL acceptance was estimated by
varying the CAL energy scale by $\pm$3\%, leading to an effect of 
$^{+0.006}_{-0.008}$. The 35m-tagger energy scale was varied by $\pm$3\%
and led to a small effect.
The measured elastic cross sections~\cite{rho,omega,phi}, the fraction of the 
photon-dissociative processes~\cite{pdisselasticratio},
and the elastic to proton-dissociative cross-section ratio~\cite{gdissfrac}
were also varied by  one standard deviation in the fitting procedure. 
The uncertainty due to the 
measured  photon-dissociative cross section was $^{+0.008}_{-0.010}$; the 
others gave small effects. The
uncertainty in modeling the non-diffractive hadronic final state was determined
from a comparison of the PYTHIA and HERWIG MC simulations. This led to a 
+0.019 uncertainty. Several strategies were employed to fit the 
experimental distributions by varying the fractions of the processes shown
in Table~\ref{table:acc}. A fit yielding the fractions of elastic, 
photon dissociation and hard and soft non-diffraction was performed.
The resulting elastic and photon-dissociative
fractions were found to be consistent 
with the ZEUS measurements, extrapolated 
to the $W_{\gamma p}$ of the present measurement. 
Global fits were made simultaneously to a number of experimental 
distributions or separately to individual distributions; 
the resulting systematic uncertainties on the CAL acceptance were negligible.
Adding all of the above contributions in
quadrature results in a CAL acceptance of $0.781^{+0.022}_{-0.016}$.

The sources of largest systematic uncertainty in the 35m-tagger acceptance
are: the uncertainty in modeling the trajectories 
of the scattered positrons, i.e., the $X$-vertex position ($\pm$0.027); 
the geometric description of the positron-beam angular spread ($\pm$0.021);
the details of the HERA beamline simulation ($\pm$0.020);
the uncertainty on the energy calibration of the photon detector ($\pm$0.015), 
and its energy nonlinearity ($\pm$0.020). All of the contributions,
of which the above are the most important, are added in
quadrature and result in an uncertainty on the 35m-tagger acceptance
of $\pm$0.050.

The 35m-tagger acceptance calculation was checked using an 
event sample containing two or more jets in the CAL, 
which was taken with a trigger independent of the 35m tagger.  
For these events, the inelasticity
 $y$ can be determined from CAL energies only.
The probability for these events to have a positron 
detected in the 35m tagger was compared to the probability
determined from the tuned photoproduction MC simulation;
the simulated events and data were  consistent within 
the quoted systematic uncertainty.

\section{Results}

The measured photon-proton total cross section is given by
\begin{displaymath}
  \sigtot = \frac {N \cdot \Delta _{\rm corr}}
           {{\cal L}\cdot{f_{\gamma}} \cdot {A_{\rm 35m}}\cdot{A_{\rm CAL}}}, 
\nonumber
\end{displaymath}
where $N$ is the number of events passing the selection cuts ($22533\pm162$),
$\cal L= $ is the integrated luminosity ($49.26\pm0.54$ nb$^{-1}$),
$f_{\gamma}$ is the photon flux factor (0.004916),
$A_{\rm 35m}$ is the 35m-tagger acceptance ($0.693\pm0.050$), and
$A_{\rm CAL}$ is the CAL acceptance ($0.781^{+0.022}_{-0.016}$).
The correction factor, $\Delta_{\rm corr}$, 
is the product of the radiative correction
to the electroweak Born-level cross section
(0.981 $\pm$ 0.007), the correction for
bremsstrahlung  background events in which the bremsstrahlung
 photon was lost due 
to the photon-tagger acceptance (0.99 $\pm$ 0.01), 
and the correction for photoproduction
events lost due to an accidental overlay with bremsstrahlung events
(1.043 $\pm$ 0.002).
The 35m-tagger acceptance and CAL acceptance
were assumed to be independent of each other.
All acceptances and correction factors were calculated for the 
$195<W_{\gamma p}<225$ GeV range of this measurement.

The photoproduction total cross section, measured at the
average photon-proton center-of-mass energy of 209 GeV, is
\begin{displaymath}
\sigtot 
 = 174 \pm 1 ({\rm stat.}) \pm 13 ({\rm syst.}) \ \mu {\rm b}. \nonumber
\end{displaymath}

\section{Discussion of results}
The photon-proton total cross section as a function
of the center-of-mass energy is shown in 
Fig.~\ref{fig:result}.
The present result is in good agreement with a measurement
from H1~\cite{h1new} at a similar
center-of-mass energy and is consistent with the previous ZEUS 
measurements~\cite{zeusold}, which it supersedes. The low-energy 
data~\cite{lowenergy}
are also shown in Fig.~\ref{fig:result}. The present result
can also be compared with an earlier ZEUS measurement~\cite{bpc1} 
of the inclusive
electroproduction cross section in the range 
$0.11 \le Q^2 \le 0.65 \,  {\rm GeV}^2$.
Extrapolating the cross section to $Q^2 = 0$, using the generalized VDM, 
yields the photoproduction total cross section. 
This is much  more model dependent and leads to a cross section of
 $187 \pm 5({\rm stat.}) \pm 14({\rm syst.}) \ \mu$b 
at a center-of-mass energy of $W_{\gamma p}=212$ GeV, 
in agreement  within errors with the present measurement.

Fits of hadronic total cross sections and an investigation of their
universal high-energy behavior have been carried out by Donnachie
and Landshoff~\cite{dl92} using the form of Eq. (\ref{eq:regge}).
A similar fit has been performed by Cudell {\it et al.}~\cite{cu00}
based on more recent hadronic data.
A ZEUS fit of the form 
\begin{eqnarray}
\sigma_{tot} = A\cdot W_{\gamma p}^{2\epsilon} 
        + B\cdot W_{\gamma p}^{-2\eta},
\label{eq:regge1}
\end{eqnarray}
where $W_{\gamma p}$ is in GeV,
to the existing
$\gamma p$ data~\cite{lowenergy,h1new} and including the present measurement
has been performed and is shown as the solid curve in Fig.~\ref{fig:result}.
The present fit has been restricted to $W_{\gamma p}>4 \ {\rm GeV}$
and the Reggeon intercept ($\alpha_{\reggeon}(0)=1-\eta$) 
has been 
fixed to the value found by Cudell {\it et al.}, $\eta =  0.358 \pm 0.015$.
The resulting fit parameters are:
\begin{eqnarray}
A = 57 \pm 5 \, \mu {\rm b}; \ B = 121 \pm 13 \, \mu {\rm b}; 
   \ \epsilon = 0.100 \pm 0.012 \, .
\label{eq:fitresult}
\end{eqnarray}
The resulting value of $\epsilon$,
related to the Pomeron intercept ($\alpha_{\pomeron}(0)=1+\epsilon$), 
is in good agreement with 
$\epsilon = 0.093 \pm 0.002$ found by Cudell {\it et al.}, a value derived
primarily from $pp$ and $\bar{p}p$ data.

A fit including a soft- and hard-Pomeron trajectory by Donnachie
and Landshoff (DL98) ~\cite{dl98} also agrees with the present
measurement within uncertainties, as shown by the dot-dashed curve in 
Fig.~\ref{fig:result}.
Other models~\cite{fs,bl99} based on the existing hadron-hadron
total cross-section data
are also in agreement with this measurement.

The optical theorem and VDM provide 
a connection between
the photon-proton total cross section and the forward elastic 
scattering amplitudes
for photoproduction of vector-meson states, $V$, via
\begin{eqnarray*}
\sigtot = \sqrt {16 \pi \cdot 
  \left. \frac{\textstyle d \sigma ^{\gamma p \to \gamma p}}
              {\textstyle  d t}             \right|_{t=0} }
 = \sum_{V=\rho^0,\omega,\phi} 
   \sqrt{16 \pi \cdot \frac{4 \pi \alpha}{f^2_V} \cdot 
         \left. \frac{\textstyle d \sigma ^{\gamma p \to Vp}}
                     {\textstyle d t}                      \right|_{t=0} } \, .
\end{eqnarray*}
The forward elastic scattering amplitudes for the 
$\gamma p \to \gamma p$ and $\gamma p \to  V p$ cross sections  
have been assumed to
be purely imaginary and $f_V$ are the photon to vector-meson coupling
constants. 

Summing only over the light vector-meson states $\rho^0$, $\omega$ and $\phi$,
using the exclusive photoproduction differential cross sections from ZEUS 
measurements~\cite{rho,omega,phi}, 
and values of $f^2_V /4 \pi$ = 2.20, 23.6 and 18.4~\cite{bauer} for
$\rho^0$, $\omega$ and $\phi$, respectively,
a value of $ 111 \pm 13 ({\rm exp.})\, \mu {\rm b}$ is obtained
for the photon-proton total cross section at $W_{\gamma p} = 70$ GeV. 
The $\rho ^0$ meson contributes about 85\% of this value.
The photon-proton total cross section at a
center-of-mass energy of $W_{\gamma p} = 70$ GeV,
obtained by interpolation between the present measurement 
and the lower energy measurements using the fit described
by Eqs.~(\ref{eq:regge1}) and (\ref{eq:fitresult}), is 
$139\pm 4$ $\mu {\rm b}$. Given the additional theoretical uncertainties
in the VDM calculation of 111 $\mu {\rm b}$, these
results are consistent.

The present measurement can also be used to test
factorization, which connects  $\gamma \gamma$, 
$\gamma p$ and $p p$ total cross sections according to
\begin{equation}
\sigma^{\gamma \gamma}_{\rm tot} \cdot \sigma^{pp}_{\rm tot} =
 (\sigma^{\gamma p}_{\rm tot})^2 \, .
\label{eq:factorization}
\end{equation} 
The $\gamma \gamma$ total
cross section has been measured at high energies by the OPAL~\cite{opal}
and L3~\cite{l3} collaborations. Using Eq.~(\ref{eq:factorization})
at the energy of the OPAL and L3 measurements requires an interpolation
from the present measurement to lower center-of-mass energies. 
Using the fit described by Eqs.~(\ref{eq:regge1}) and (\ref{eq:fitresult})
and the Cudell {\it et al.} parameterization of the $pp$ total cross
sections, a value $\sigma^{\gamma \gamma}_{\rm tot}$
= $436 \pm 28$ nb ($468 \pm 30$ nb) is obtained for
a center-of-mass energy of 68 (95) GeV, in agreement with the OPAL
measurement of $439^{+45}_{-41}$ ($464^{+76}_{-62}$) nb at 
those energies. Measurements from L3 agree
 within uncertainties with those from OPAL for $W_{\gamma \gamma}<100$ GeV.
Thus the present $\sigtot$ measurement is consistent with
the factorization hypothesis of Eq.~(\ref{eq:factorization}).

At $W_{\gamma \gamma}$ = 120.4 GeV, L3~\cite{l3} finds 
   $572.0 \pm 3.3 ({\rm stat.}) \pm 53 ({\rm exp.~syst.}) 
          \pm 89 ({\rm MC~syst.}) $ nb, while at an energy of 158.7 GeV, they
find $734.1 \pm 8.7 ({\rm stat.}) \pm 102 ({\rm exp.~syst.}) 
          \pm 202 ({\rm MC~syst.})$ nb. There is an additional 
uncertainty of $\pm 5\%$ due to the luminosity measurement.
Interpolations from the present $\sigtot$
measurement give $491 \pm 35$ nb and $521 \pm 43$ nb, respectively.
Within the large systematic uncertainties of the L3 measurement, 
these data are also consistent with factorization.

In earlier parameterizations~\cite{dl92,cu00,dl98,fs}, the high-energy dependence
has effectively been determined from
the $pp$ and $\bar{p}p$ data.
The compatibility of the current ZEUS $\sigtot$ measurement with
this high-energy behavior indicates a universality of the
energy dependence of the photon-proton total cross section with respect
to that of hadron-proton total cross sections. 

\section*{Acknowledgements}

We thank the DESY directorate for their strong support and encouragement.
The special efforts of the HERA machine group in the collection of the data 
used in this paper are gratefully acknowledged.
We thank the DESY computing and 
network services for their support.  
The design, construction and installation of the ZEUS detector 
have been made possible by the ingenuity and effort of many people 
from DESY and home institutes who are not listed as authors.
We are grateful for the helpful discussions and correspondence with
A.~Arbuzov, D.~Bardin, M.~Block, H.~Spiesberger and T.~Sj\"{o}strand.

\newpage



\newpage

\begin{figure}[htbp!]
\epsfysize=18cm
\centerline{\epsffile{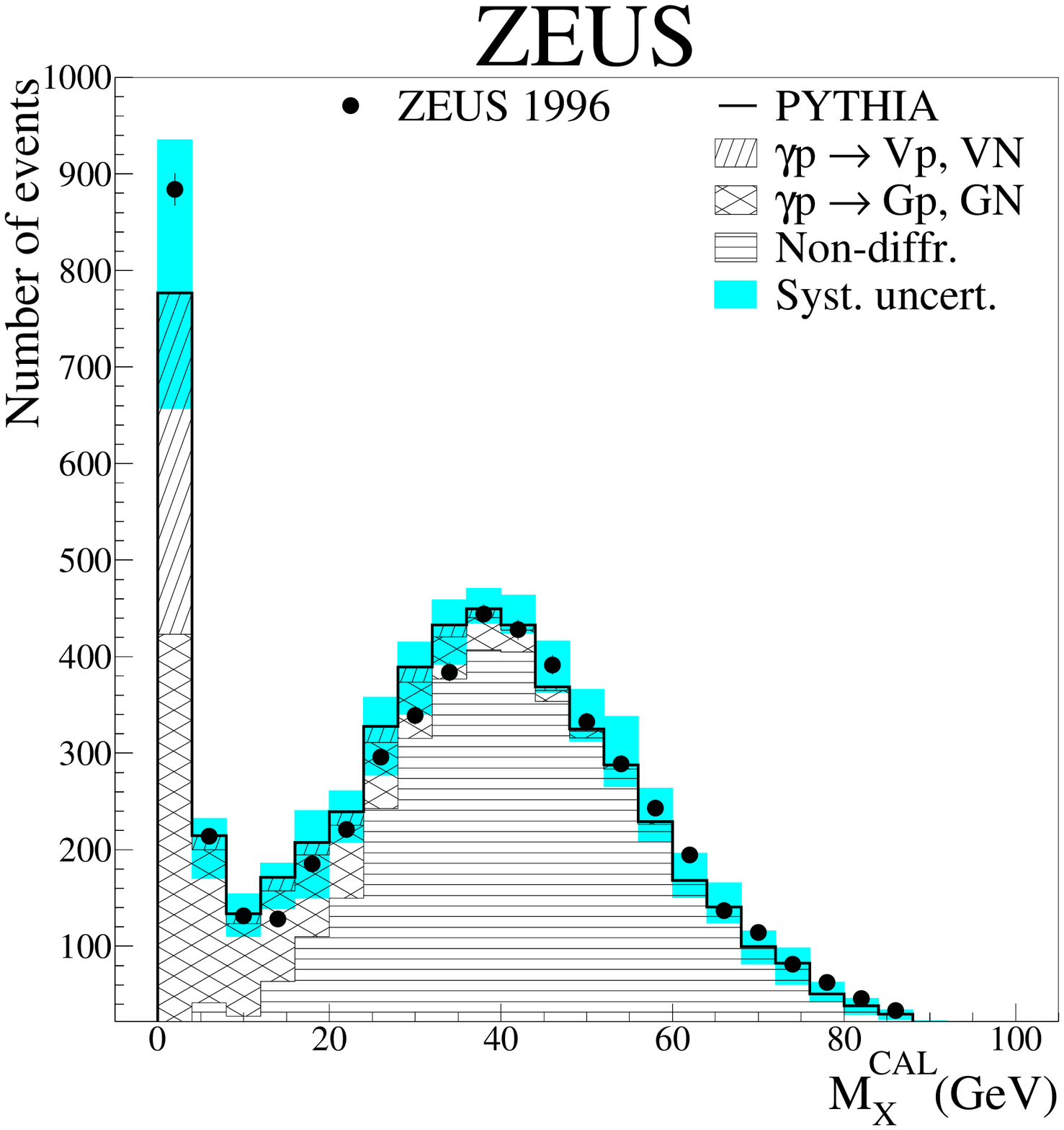}}
\caption{$M_X^{\rm CAL}$ distribution for data (filled circles) and fitted 
total photoproduction MC sample (histogram with systematic uncertainty band).
The fit is made to the $M_X^{\rm CAL}$ and $N_{\rm cells}$ distributions. 
Cumulative subprocess contributions are also shown. The elastic and 
proton-dissociative samples have been combined, as have the
photon-dissociative and double-dissociative 
samples and the soft and hard non-diffractive samples.
} 
\label{fig:fig2mxlin}
\end{figure}

\clearpage

\begin{figure}[htbp!]
\epsfysize=18cm
\centerline{\epsffile{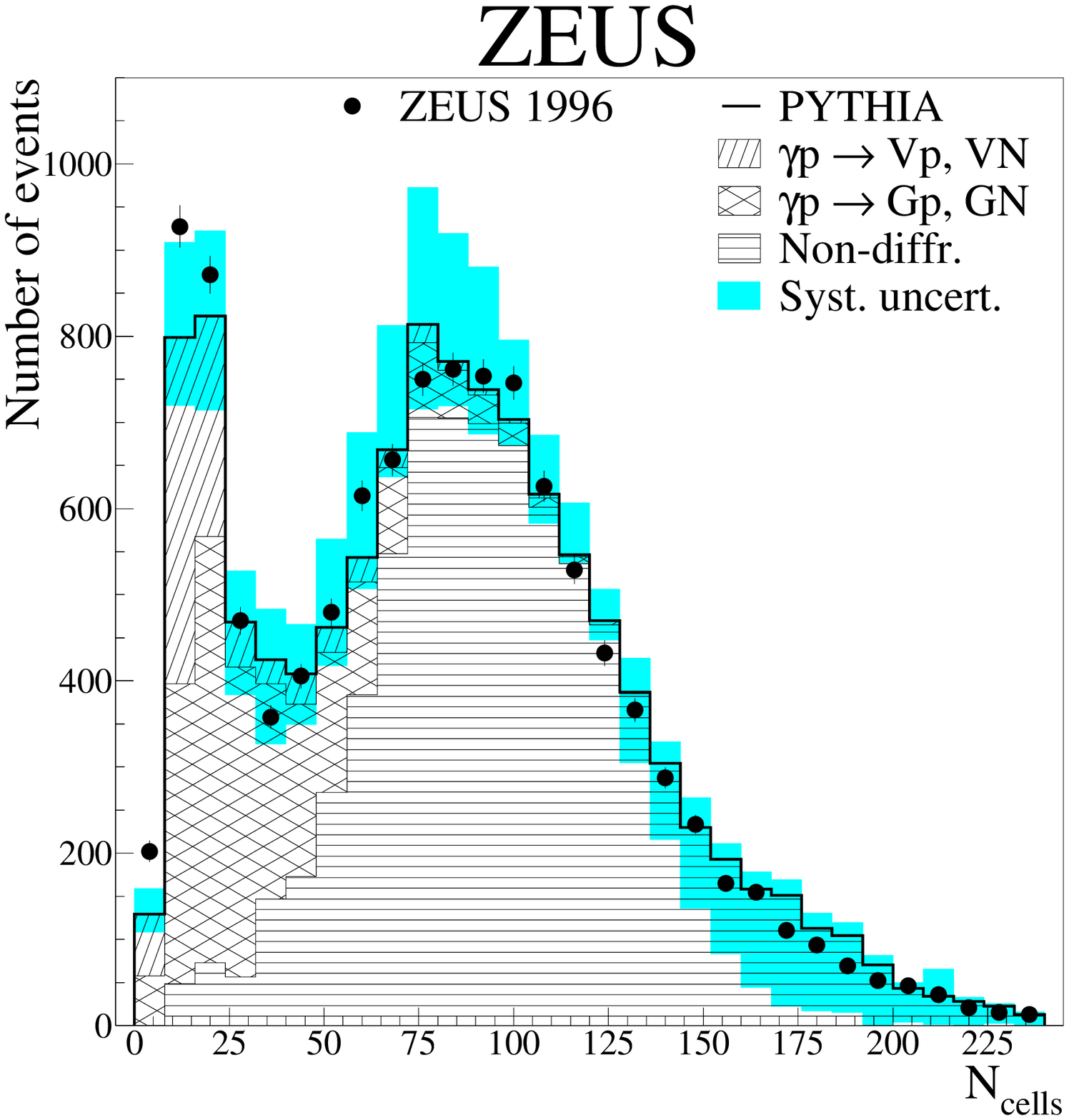}}
\caption{$N_{\rm cells}$ distribution for data (filled circles) and fitted 
total photoproduction MC sample (histogram with systematic uncertainty band).
The fit is made to the $M_X^{\rm CAL}$ and $N_{\rm cells}$ distributions. 
Cumulative subprocess contributions are also shown. The elastic and 
proton-dissociative samples have been combined, as have the
photon-dissociative and double-dissociative 
samples and the soft and hard non-diffractive samples.
}
\label{fig:fig2ncellin}
\end{figure}

\clearpage

\begin{figure}[htbp!]
\epsfysize=18cm
\centerline{\epsffile{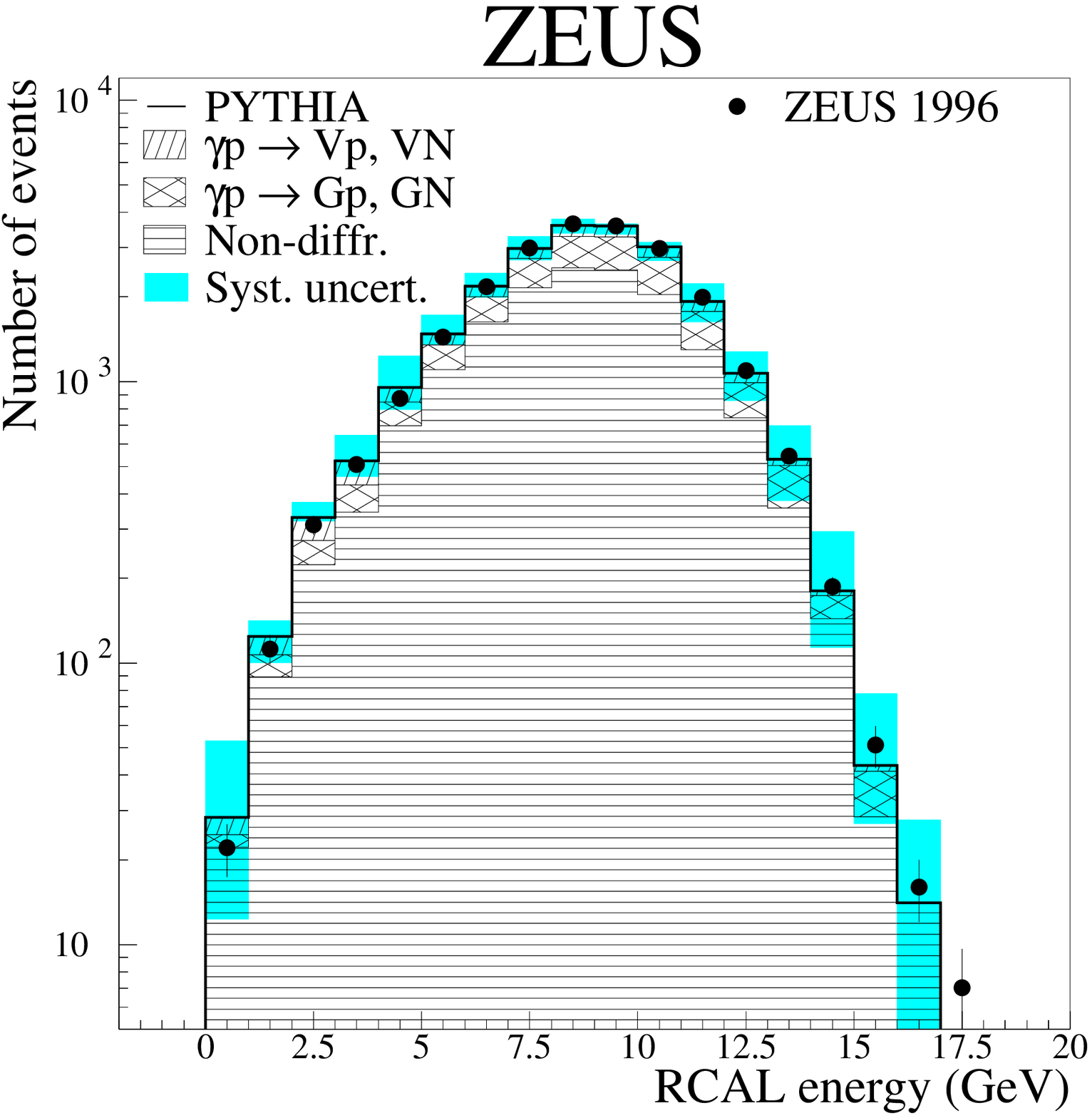}}
\caption{RCAL energy distribution for data (filled circles) and fitted total
photoproduction MC sample (histogram with systematic uncertainty band).
The fit is made to the $M_X^{\rm CAL}$ and $N_{\rm cells}$ distributions. 
Cumulative subprocess contributions are also shown. The elastic and 
proton-dissociative samples have been combined, as have the
photon-dissociative and double-dissociative 
samples and the soft and hard non-diffractive samples.
}
\label{fig:fig2rcallog}
\end{figure}

\clearpage
\begin{figure}[htbp!]
\epsfysize=18cm
\centerline{\epsffile{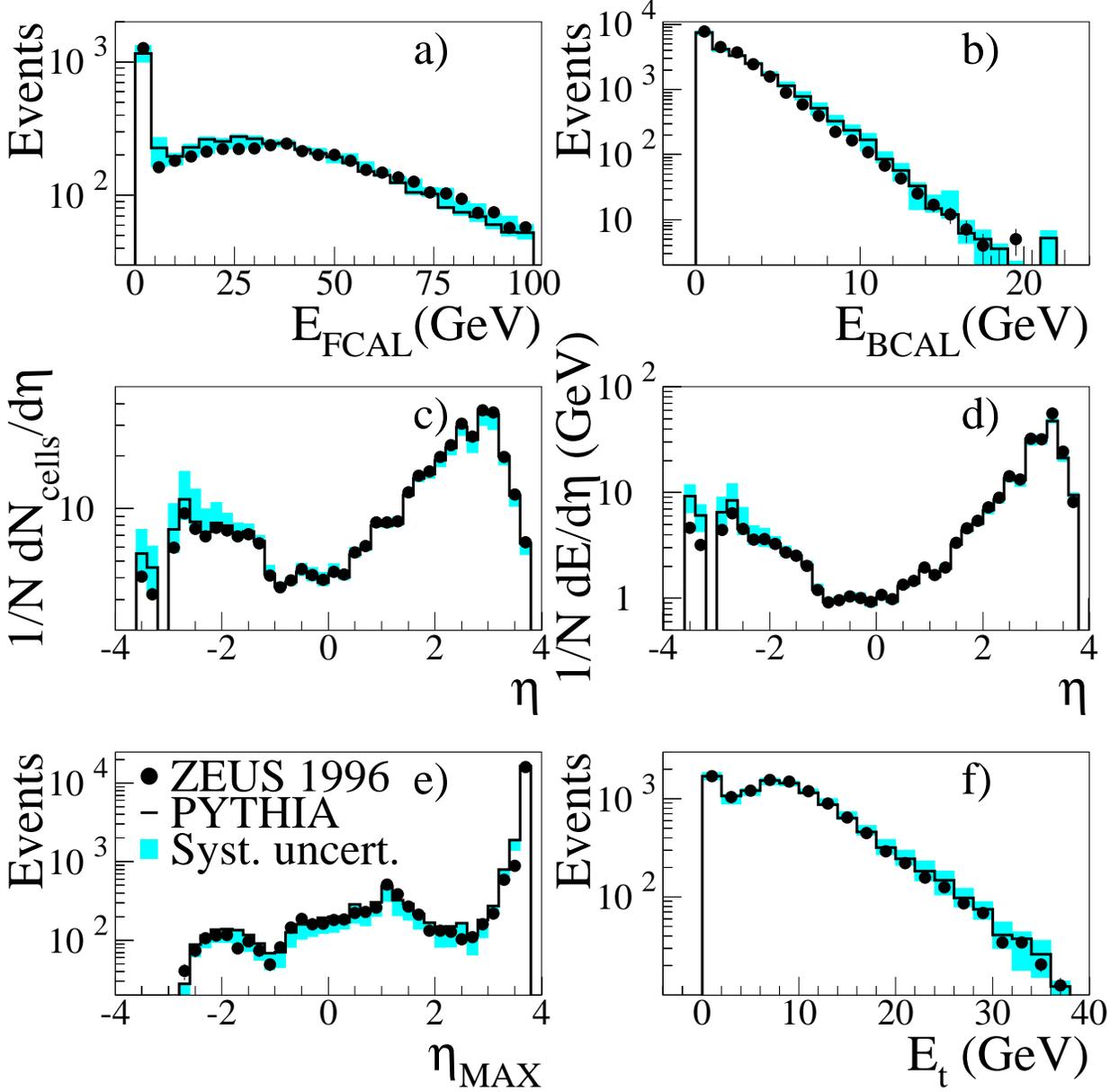}}
\caption{Comparison of the data (filled circles) and fitted 
photoproduction MC sample (histogram with systematic uncertainty band) for
a) FCAL energy, b) BCAL energy, c) pseudorapidity of the CAL cells, 
d) energy-weighted pseudorapidity of the CAL cells, e) pseudorapidity of 
the most forward energy deposit of those CAL EMC cells above 80 MeV or the
CAL HAC cells above 140 MeV
and f) total transverse energy of each event. 
The fit is made to the $M_X^{\rm CAL}$ and $N_{\rm cells}$ distributions. 
}
\label{fig:fig2rest}
\end{figure}

\clearpage

\begin{figure}[htbp!]
\epsfysize=18cm
\centerline{\epsffile{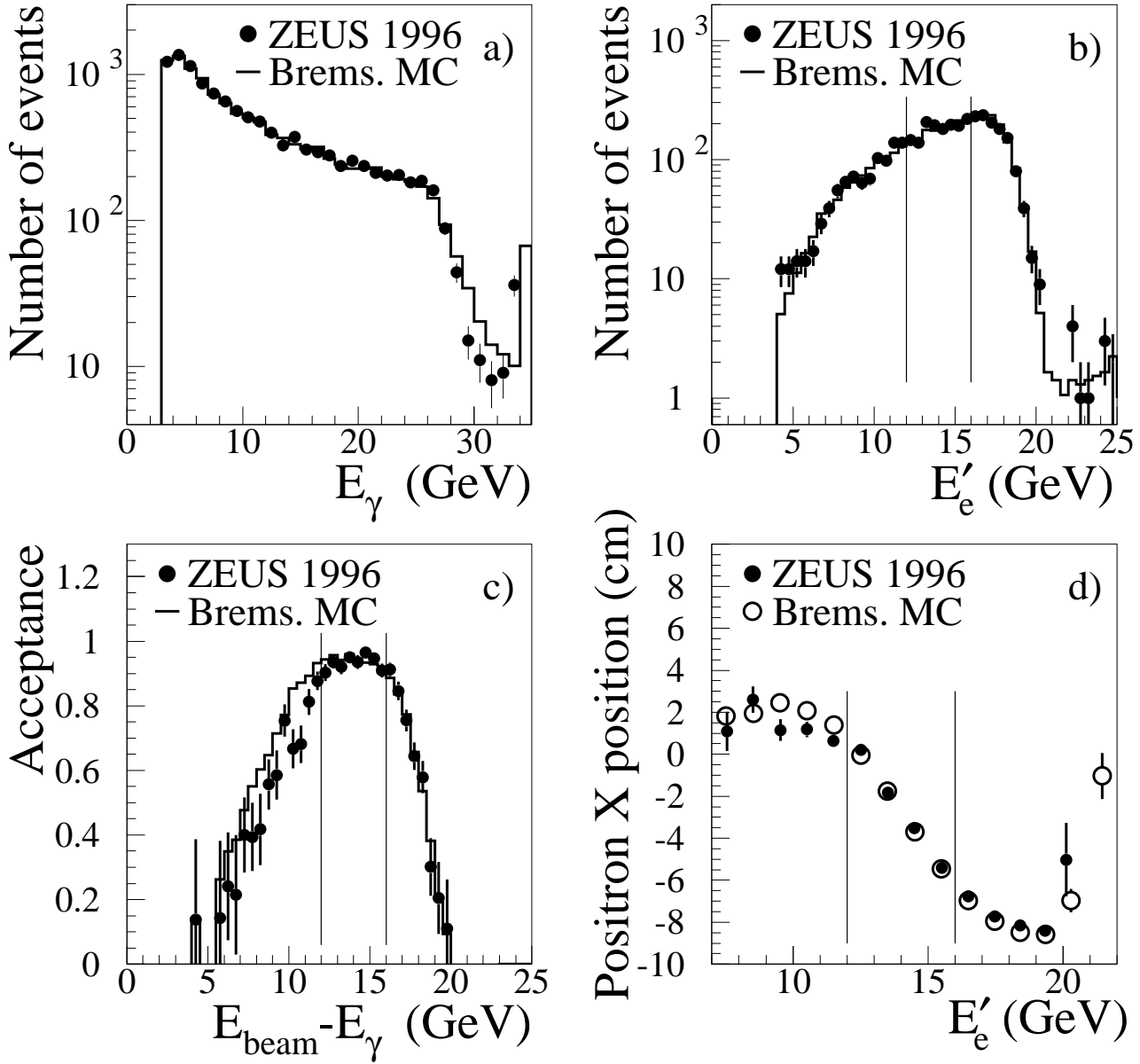}}
\caption{Distributions for the bremsstrahlung data (filled circles):
a) photon energy, b) positron energy, c) 35m-tagger bremsstrahlung
acceptance as a function of the predicted positron energy and d) positron
position vs. positron energy. In (b-d), the region used for the scattered 
positron energy in the $\sigtot$ measurement, $12<E'_e<16$ GeV, is shown 
by the vertical lines. The tuned bremsstrahlung  Monte
Carlo simulation is shown in (a-c) as the histogram and in d) as the open
circles.
}
\label{fig:bsopt}
\end{figure}

\begin{figure}
\centerline{
   \epsfysize 18 truecm \epsfbox{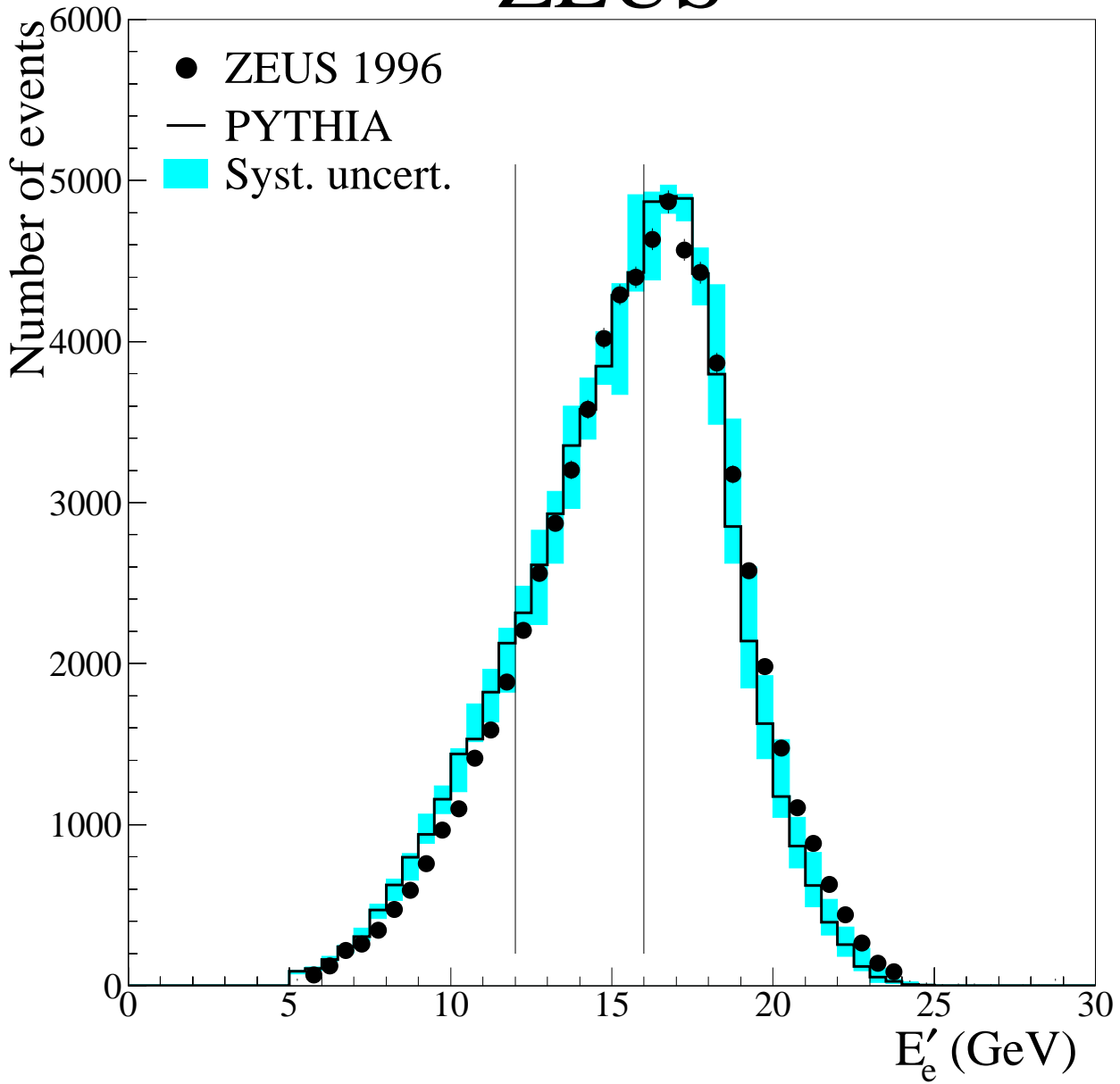}}
\caption{The positron energy distribution for photoproduction
    data (filled circles)
    and tuned  MC events (histogram with
    systematic uncertainty band).
    The selected region, $12<E'_e<16$ GeV, is shown by the vertical lines.
}
 \label{fig:gpopt}
\end{figure}

\begin{figure}
\centerline{   \epsfysize 18 truecm \epsfbox{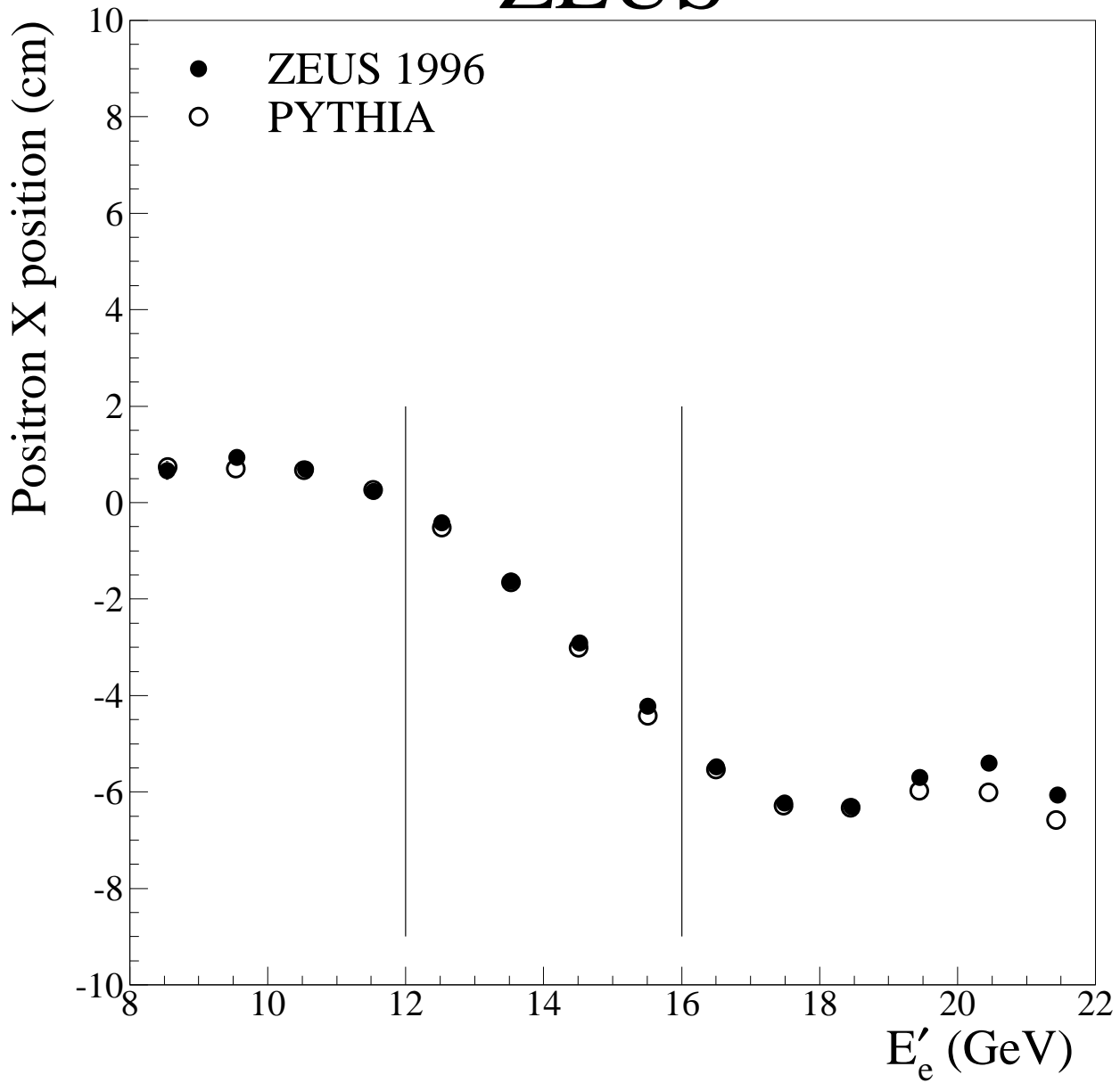}}
\caption{The correlation between the position and energy of the scattered
  positrons for photoproduction data (filled circles) and MC events 
  after tuning of the $X$ tilt and $X$ position (open circles).
  The selected region, $12<E'_e<16$ GeV, is shown by the vertical lines.
 \label{fig:gpcorr}}
\end{figure}

\begin{figure}
\centerline{
   \epsfysize 18 truecm \epsfbox{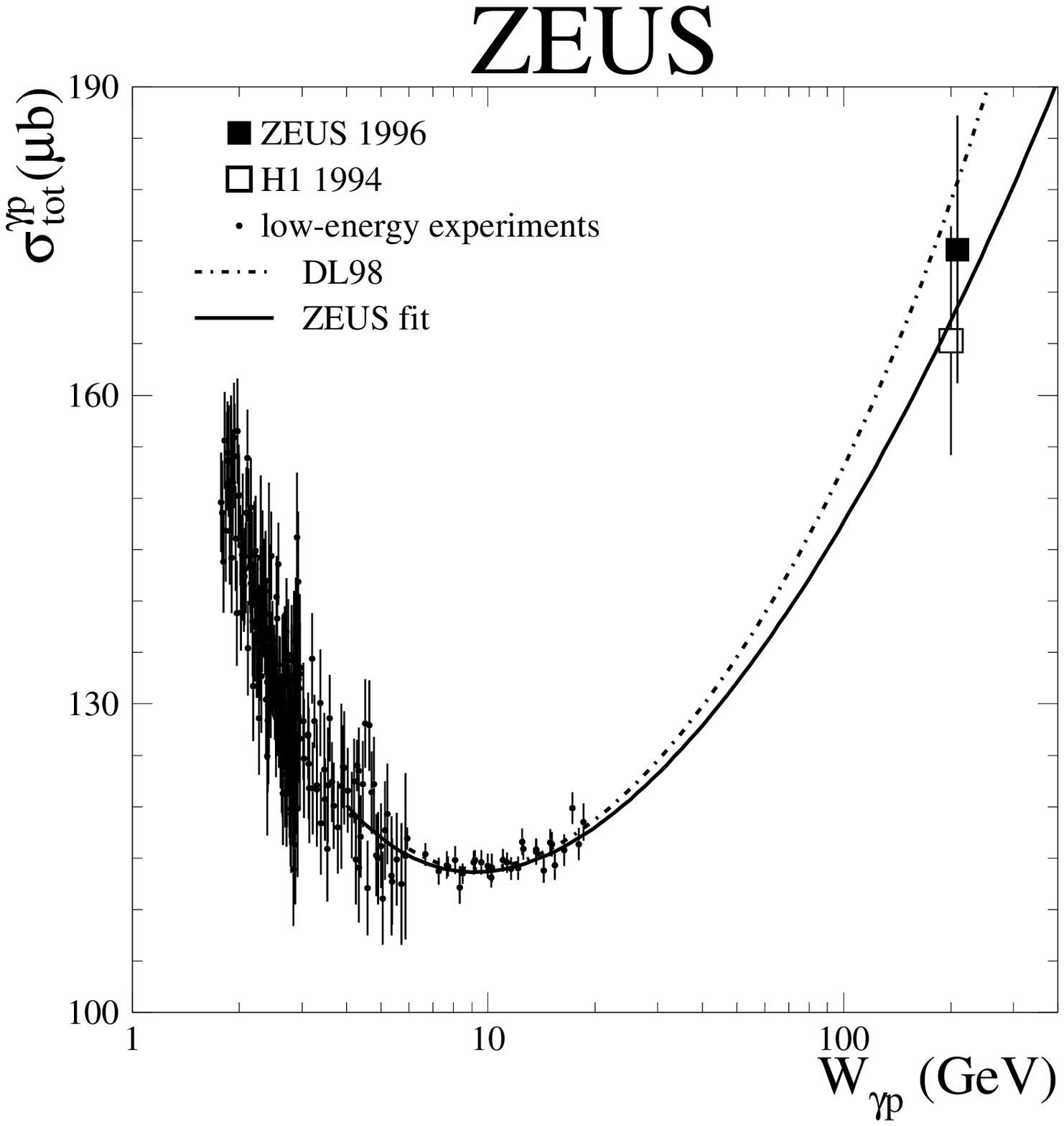}}
\caption{The photon-proton total cross section as a function
of photon-proton center-of-mass energy.
The present measurement is shown as the filled square.
Also shown are the published H1 value (open square), 
the low-energy data (filled circles),
the DL98 parameterization (dot-dashed curve)
and the ZEUS fit (solid curve)
described by Eqs.~(\ref{eq:regge1}) and~(\ref{eq:fitresult}), see text.
 \label{fig:result}}
\end{figure}


\begin{thebibliography}{999}
%
\bibitem{regge}
  P.D.B.~Collins, {\em An Introduction to Regge Theory and High Energy 
   Physics}, Cambridge University Press (1977).
%
\bibitem{dl92}
  A.~Donnachie and P.V.~Landshoff, Phys.~Lett. {\bf B296}, 227 (1992).
%
\bibitem{cu00}
  J.R.~Cudell {\it et al.}, Phys.~Rev. {\bf D61}, 034019 (2000), 
   Erratum-ibid.  {\bf D63}, 059901 (2001).
%
\bibitem{dl98}
  A.~Donnachie and P.V.~Landshoff, Phys.~Lett. {\bf B437}, 408 (1998).
%
\bibitem{fs}
  C.~Friberg and T.~Sj\"{o}strand, JHEP {\bf 09}, 010 (2000), 
  and references therein.
%
\bibitem{vmd}
  J.J.~Sakurai, Ann.~Phys.~(NY) {\bf 11}, 1 (1960);\\
  J.J.~Sakurai, Phys.~Rev.~Lett. {\bf 22}, 981 (1969).
%
\bibitem{bauer}
  T.H.~Bauer {\it et al.}, Rev.~Mod.~Phys. {\bf 50}, 261 (1978), 
      Erratum-ibid. {\bf 51}, 407 (1979).
%
\bibitem{lowenergy}
  D.O.~Caldwell {\it et al.}, Phys.~Rev.~Lett.~{\bf 40}, 1222 (1978);\\
  S.I.~Alekhin {\it et al.}, CERN Report HERA 87-01 (1987).
%

%
\bibitem{zeusold}
  ZEUS Collab., M.~Derrick {\it et al.},
    Phys.~Lett. {\bf B293}, 465 (1992);\\
  ZEUS Collab., M.~Derrick {\it et al.},
    Z.~Phys. {\bf C63}, 391 (1994).
%
%
%
\bibitem{rho}
  ZEUS Collab., M.~Derrick {\it et al.},
    Z.~Phys. {\bf C69}, 39 (1995).       
%
\bibitem{omega}
  ZEUS Collab., M.~Derrick {\it et al.},
    Z.~Phys. {\bf C73}, 73 (1996).          
%
\bibitem{phi}
  ZEUS Collab., M.~Derrick {\it et al.},
    Phys.~Lett. {\bf B377}, 259 (1996).   
%
\bibitem{pdisselasticratio}
  ZEUS Collab., J.~Breitweg {\it et al.},
    Eur.~Phys.~J. {\bf C2}, 247 (1998);\\   
  K.~Desler, Ph.D.~Thesis, Universit\"{a}t Hamburg  (2000), (unpublished).
%
\bibitem{gdissfrac}
  ZEUS Collab., J.~Breitweg {\it et al.},
    Z.~Phys. {\bf C75}, 421 (1997).      
%
\bibitem{h1new}
 H1 Collab., S.~Aid {\it et al.}, Z.~Phys. {\bf C69}, 27 (1995).   
%
\bibitem{gribov}
    V.N.~Gribov {\it et al.}, Sov.~Phys.~JETP {\bf 14}, 1308 (1962).
%
\bibitem{lohrmann}
     B.~Bade{\l}ek, J.~Kwieci\'{n}ski, and A.~Sta\'{s}to, 
       Z.~Phys. {\bf C74}, 297 (1997);\\
     D.~Schildknecht and H.~Spiesberger, BI-TP 97/25, {\em hep-ph/9707447}; \\
     D.~Schildknecht, Acta Phys. Polon. {\bf B28}, 2453 (1997). 
%
\bibitem{zeusdet}
  ZEUS Collab., U.~Holm (ed.), {\it The ZEUS Detector,} Status Report,
       (unpublished), DESY, 1993;\\
  {\tt http://www-zeus.desy.de/bluebook/bluebook.html}~.
%
\bibitem{cal}
  M.~Derrick {\it et al.}, Nucl.~Instr. and~Meth. {\bf A309}, 77 (1991);\\
  A.~Andresen {\it et al.}, Nucl.~Instr. and~Meth. {\bf A309}, 101 (1991);\\
  A.~Caldwell {\it et al.}, Nucl.~Instr. and~Meth. {\bf A321}, 356 (1992);\\
  A.~Bernstein {\it et al.}, Nucl.~Instr. and~Meth. {\bf A336}, 23 (1993).
%
\bibitem{ctd}
  N.~Harnew {\it et al.}, Nucl.~Instr. and~Meth. {\bf A279}, 290 (1989);\\
  B.~Foster {\it et al.}, Nucl.~Phys.~Proc.~Suppl. {\bf B32}, 181 (1993);\\
  B.~Foster {\it et al.}, Nucl.~Instr. and~Meth. {\bf A338}, 254 (1994).
%
\bibitem{lumi}
  D.~Kisielewska {\it et al.}, Nukleonika {\bf 31}, 205 (1986);\\
  J.~Andruszk\'{o}w {\it et al.}, DESY Report 92-066 (1992); \\
  J.~Andruszk\'{o}w {\it et al.}, DESY Report 01-041 (2001).
%
\bibitem{caltrigger}
  W.H.~Smith {\it et al.}, Nucl.~Instr. and~Meth. {\bf A355}, 278 (1995).
%
\bibitem{fastclear}
  B.G.~Bylsma {\it et al.}, Nucl.~Instr. and~Meth. {\bf A337}, 512 (1994).
%
\bibitem{pythia}
  T.~Sj\"{o}strand, PYTHIA 5.7, {\it hep-ph/9508391};\\
  T.~Sj\"{o}strand, Comp.~Phys.~Comm.~{\bf 82}, 74 (1994).
%
\bibitem{heracles}
   A. Kwiatkowski, H. Spiesberger  and  H.-J. M\"ohring,
      Comp.~Phys.~Comm.~{\bf 69}, 155 (1992);\\
   H. Spiesberger, HERACLES 4.6 Manual,
      {\tt http://www.desy.de/$\sim$hspiesb/heracles.html}~.
%
%
\bibitem{herwig}
  G.~Marchesini {\it et al.}, HERWIG 5.9, {\it hep-ph/9607393};\\
  G.~Marchesini {\it et al.}, Comp.~Phys.~Comm.~{\bf 67}, 465 (1992).
%
\bibitem{cteq} 
  CTEQ Collab., H.L.~Lai {\it et al.}, Phys.~Rev. {\bf D55}, 1280 (1997).
%
\bibitem{grv}
 M.~Gl\"{u}ck, E.~Reya and A.~Vogt, Phys.~Rev. {\bf D45}, 3986 (1992);\\
 M.~Gl\"{u}ck, E.~Reya and A.~Vogt, Phys.~Rev. {\bf D46}, 1973 (1992).
%
\bibitem{bremmc} 
  K.~Piotrzkowski and L.~Suszycki, 
   {\it Proceedings of the Workshop on Physics at HERA,
  Oct. 1991}, Volume 3,
  W.~Buchm\"{u}ller and G.~Ingelman (eds.), DESY (1992), p. 1463.
%
\bibitem{hector}
  A.~Arbuzov {\it et al.}, Comp.~Phys.~Comm.~{\bf 94}, 128 (1996); \\
  A.~Arbuzov {\it et al.}, Hector 1.00 Manual, DESY Report 95-185,
   {\it hep-ph/9511434}.
%
%
\bibitem{bpc1}
  ZEUS Collab., J.~Breitweg {\it et al.}, Eur.~Phys.~J. {\bf C7}, 609 (1999).
%
%
\bibitem{bl99}
  M.M.~Block {\it et al.}, Phys.~Rev. {\bf D60}, 054024 (1999);\\
  M.M.~Block, F.~Halzen and T.~Stanev, Phys.~Rev. {\bf D62}, 077501 (2000).
%
\bibitem{opal}
 OPAL Collab., G.~Abbiendi {\it et al.}, Eur.~Phys.~J. {\bf C14}, 199 (2000).
%
%
\bibitem{l3} 
 L3 Collab., M. Acciarri {\it et al.}, Phys.~Lett. {\bf B519}, 33 (2001).

%
%
%
%
\end{thebibliography}
\end{document}